\documentclass{article}

\usepackage{arxiv}

\usepackage[utf8]{inputenc} 
\usepackage[T1]{fontenc}    
\usepackage{hyperref}       
\usepackage{url}            
\usepackage{booktabs}       
\usepackage{amsfonts}   
\usepackage{amsmath}
\usepackage{nicefrac}       
\usepackage{microtype}      
\usepackage{lipsum}
\usepackage{graphicx}
\usepackage{footnote}
\usepackage{soul}
\usepackage{color}
\usepackage{soul}
\usepackage{float}
\usepackage{subcaption}

\title{Cloudy with a Chance of Anomalies: Dynamic Graph Neural Network for Early  Detection of Cloud Services' User Anomalies}

\author{Revital Marbel\\
School of Computer Science\\ Holon Institute of Technology (HIT), Israel\\
  \texttt{marbelr@hit.ac.il}\\
 \And Yanir Cohen \\Department of Computer Science\\ Ariel Cyber Innovation Center.\\
  Ariel University, Kiryat Hamada 3, Ariel\\
      \And  Ran Dubin\\Department of Computer Science\\ Ariel Cyber Innovation Center.\\
  Ariel University, Kiryat Hamada 3, Ariel\\
          \And Amit Dvir\\Department of Computer Science\\ Ariel Cyber Innovation Center.\\
  Ariel University, Kiryat Hamada 3, Ariel\\
        \And Chen Hajaj\\Department of Industrial Engineering and Management\\Data Science and Artificial Intelligence Research Center\\ Ariel University, Israel\\
}

\begin{document}
\maketitle
\begin{abstract}
Ensuring the security of cloud environments is imperative for sustaining organizational growth and operational efficiency. As the ubiquity of cloud services continues to rise, the inevitability of cyber threats underscores the importance of preemptive detection. This paper introduces a pioneering time-based embedding approach for Cloud Services Graph-based Anomaly Detection (CS-GAD), utilizing a Graph Neural Network (GNN) to discern anomalous user behavior during interactions with cloud services.
Our method employs a dynamic tripartite graph representation to encapsulate the evolving interactions among cloud services, users, and their activities over time. Leveraging GNN models in each time frame, our approach generates a graph embedding wherein each user is assigned a score based on their historical activity, facilitating the identification of unusual behavior. Results demonstrate a notable reduction in false positive rates (2-9\%) compared to prevailing methods, coupled with a commendable true positive rate (100\%). The contributions of this work encompass early detection capabilities, a low false positive rate, an innovative tripartite graph representation incorporating action types, the introduction of a new cloud services dataset featuring various user attacks, and an open-source implementation for community collaboration in advancing cloud service security.
\end{abstract}

\section{Introduction}
Detecting anomalous behavior in cloud services is crucial for maintaining a secure environment and mitigating potential threats~~\cite{garg2020abc, nedelkoski2019anomaly}. Anomalies can manifest in various forms, including user behavior anomalies and network anomalies.
Anomalies often indicate potential security incidents, such as data breaches, unauthorized access, or malicious activities \cite{zhang2019cross}. By detecting anomalous behavior in the early stages, organizations can quickly respond to potential threats and mitigate the risks associated with these incidents. Therefore, early detection and prevention can significantly reduce the impact of security breaches and minimize the potential damage to an organization's reputation, finances, and operations.

Additionally, detecting anomaly behavior complies with data protection regulations:
Organizations are subject to various data protection regulations, various regulations exist for different purposes and may prevent an organization from trading in specific regions or dealing with large companies. Detecting and addressing user and network anomalies in cloud services is essential to maintaining compliance with these regulations. Failure to detect and respond to security incidents can result in significant fines, penalties, and reputational damage.Anomalies may also indicate unauthorized access to sensitive data, attempts to tamper with or alter data, or events that could lead to data loss or unavailability. By detecting and addressing these issues, organizations can maintain the trust of their customers and partners while ensuring the continuous operation of their cloud services.

Targeting end-users in order to infiltrate an organization's cloud service has become more and more common ~\cite{alwaheidi2022data}. 
One example of such an attack is abducting a cloud account and utilizing its computing power to mine bitcoins ~\cite{achar2022cloud}.
These attacks are hard to detect since there are thousands of API calls every minute, and figuring out that the user abuses its free access to the cloud for crypto-hijacking is not straightforward since the user has permission to perform the given action. Alas, restricting the user's permission to a minimum can slow software development since requesting and obtaining permissions is a timely process. Limiting a developer to specific permissions is currently the best practice but it creates a process in which the developer has to repeatedly request the system admin for permissions. As such, this paper provides a method for developers to sustain the system's permissions while ensuring high security. 
Several works try to tackle the problem of developing auto models for anomaly detection based on machine learning methods ~\cite{huch2018machine,zhang2019deep}, Fuzzy Time Series Forecast Model ~\cite{tran2018multivariate} and Long Short Term Memory (LSTM) ~\cite{du2017deeplog}.
Common anomaly detection methods mostly focus on detecting abnormal transactions between the user and the cloud but don't consider the user behavior over time ~\cite{garg2019hybrid,dwivedi2021gaussian} or possible connections between users that have similar behavior ~\cite{he2020spatiotemporal,zhang2019cross,garg2020abc}. Hence, these methods fail to detect slow and abnormal user activity changes.

\begin{figure}
\includegraphics[width=\linewidth]{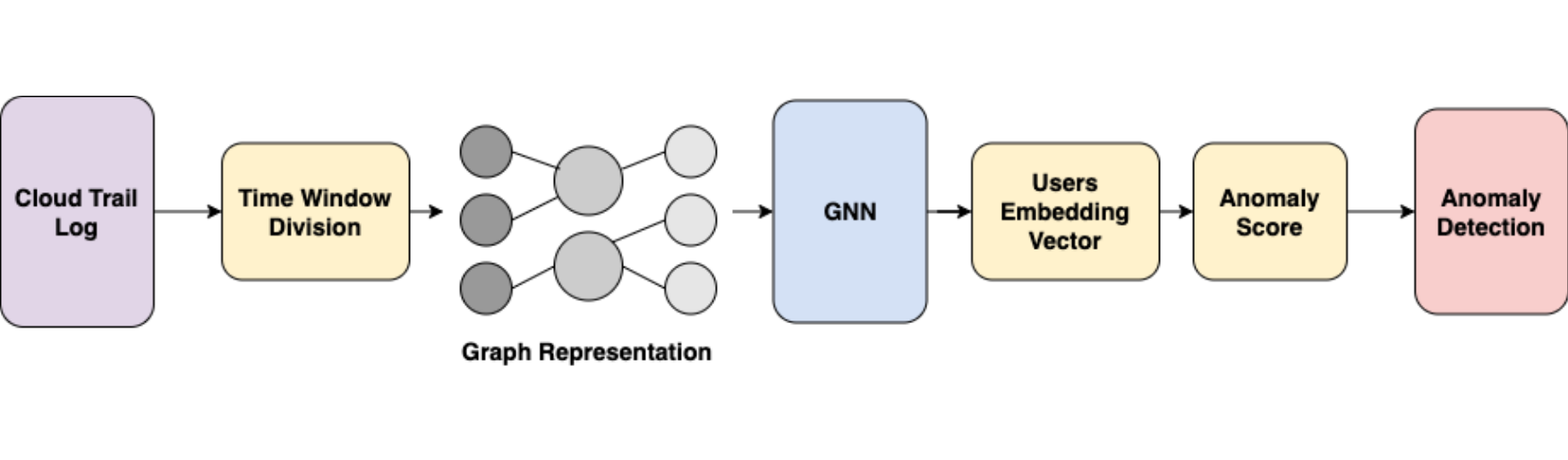}
  \caption{CS-GAD Main Flow}
  \label{fig:GraphPicture}
\end{figure}

\textbf{Our contribution}: This work presents a novel method that models users' interactions with the cloud over time and uses a Graph Neural Network-based (GNN-based) model to identify users' behavior changes. The model is based on the assumption that learning the similarity between users' behavior, when accessing a certain cloud or using a certain activity, helps detect anomalies in the early phase of the attack. 
This study approaches the detection of users and network anomaly behavior by representing the network of users, servers, and actions as a dynamic graph. Herein, a graph neural network (GNN) model is used in each time frame to create an embedding vector for each node in the graphs. This vector representation encapsulates the behavior of each component with respect to other components in the graph. The anomaly detection is determined using a custom anomaly score function, which is defined for each node in the graph.

Specifically, this paper presents the following contribution:
\begin{itemize}
\item Detecting User Anomalies in Early Stage: We present a novel method for identifying potential malicious intent or activity of cloud users. This early detection capability allows for a timely response and prevents security breaches or minimizes their impact. 
Moreover, the early detection mechanism significantly reduces the potential for large-scale damage, reinforcing the system's overall cybersecurity framework and reliability.
\item Low Rate of False Positives: Another contribution of this work lies in developing a model that boasts a significantly low rate of false positives. Conventionally, systems have struggled to maintain a balance between accurate attack recognition and limiting the occurrence of false alarms. 
Achieving a lower false positive rate ensures that security alerts retain their urgency and are not disregarded due to over-saturation. This can be useful for timely detection and response to legitimate threats, thereby reinforcing the overall security framework. 
\item  New Graph Representation with Tripartite Graph and Action Types: This research introduces a novel graph representation approach that leverages a tripartite graph. The tripartite graph consists of three distinct node sets: users, actions, and resources. In contrast to conventional graph representations, our approach incorporates action types as an additional dimension, providing an enhanced representation. Each action node represents a specific type of user activity, such as file uploads, database queries, or resource creation. By incorporating action types, we capture more details about user interactions with cloud services, enabling a wider encapsulation of user behavior.
\item New Cloud Services Dataset: This research presents a significant contribution in the form of a new cloud services dataset encompassing various real-world cloud usage scenarios. The dataset includes five types of user attacks, representing distinct forms of malicious or anomalous behavior within cloud services. It provides a comprehensive environment for evaluating and testing anomaly detection algorithms, accompanied by realistic attack scenarios and annotated ground truth labels. This dataset enables future works to effectively explore user anomaly behavior in cloud services\footnote{\url{https://github.com/yanir75/GNN-AnomalyDetection/raw/main/Datasets/attack.json}}. 
\item Open Source Implementation: In the spirit of open science, this work contributes an open-source implementation, which is publicly available and can be accessed by the community. This open-source implementation includes the code and resources for our proposed model and a comparative baseline model for detecting user anomalies in cloud services\footnote{\url{https://github.com/yanir75/GNN-AnomalyDetection.git}}.
\end{itemize}

\section{Related Work}
Security in cloud-based networks has been well-researched. Previous works include protocols for improving group data sharing \cite{ebenezer2019block}, ensuring the integrity of cloud data and the privacy of sensitive information \cite{gao2021checking},
failures detection in the cloud ~\cite{wang2021online,fu2012hybrid} or analyzing network traffic in order to identify anomalies in the traffic generated in and out of the cloud \cite{nadeem2022preventing}. Many infiltration attacks on cloud accounts are done through the end user \cite{haber2022cloud}. This is done with social engineering (tricking the user into providing information that helps infiltrate the cloud account), or by obtaining the user credentials in another way \cite{guan2022cloud}. End users in most of the cloud accounts can escalate permissions due to misconfigurations and exploits. Thus, recognizing end-user anomaly provides an additional layer of defense mechanism on top of utilizing the cloud resources, since each cloud provides a kind of record to the API calls performed by all the end users \cite{fectivecloud2}.

The world of Deep Learning (DL) keeps growing, and its use for spotting unusual patterns or anomalies is drawing a lot of attention. 
Many Deep Learning techniques were studied for detecting intrusion-based attacks on the cloud,  Thirumalairaj et al.\cite{thirumalairaj2020intelligent} suggested a Support Vector Machine (SVM) classifier to detect infiltrations. Aboueata et al. \cite{aboueata2019supervised} used an additional Artificial Neural Network (ANN) layer to detect intrusions or anomalous behavior in the cloud environment. 
Other methods used Auto Encoders for detecting DDoS attack traffic in the cloud \cite{bhardwaj2020hyperband}, Recurrent Neural Network (RNN) for detecting intrusions in SDN-based Networks \cite{tang2019intrusion} or A meta-heuristic assisted model \cite{rm2022hybrid} including K-means clustering model, centroids selection and hybrid optimization methods for cloud intrusion detection.


Adding to the strengths of DL, Graph Neural Networks (GNNs) have also shown promise in anomaly detection. Anomaly detection using graph representation is a research field covering both theoretical and AI (GNN)- based works. Theoretical works include Subgraph-based techniques ~\cite{morishima2021scalable,zhuang2023subgraph} that aim to identify malicious connections by detecting substructures within the graph. Community-based methods ~\cite{chen2012community, kopp2018community, francisquini2022community} aim to detect nodes that breach community boundaries and Bayesian approaches ~\cite{odiathevar2022bayesian, perusquia2022bayesian} utilize a learned statistical model to measure anomalies. A spectral graph embedding approach is presented by Modell et al. ~\cite{modell2021graph} using bipartite graphs and spectral graph embedding.  GNNs are particularly suited for handling structured data, such as networks or graphs ~\cite{ma2021comprehensive, jin2021anemone, ning2023mst,li2022deepag}. 



\section{Our Approach}
\label{sec:our_approach}

The problem of detecting anomalies in cloud services networks can be broken down into three main sub-problems: Graph representation of the network, designing a model for graph embedding, and defining an anomaly score function that gives a high score for abnormal behavior. The next sections detail our proposed method for every sub-problem, ending with a whole solution to the problem of user behavior anomaly detection.

\subsection{Graph Representation}
Inspired by the work of ~\cite{modell2021graph}, which constructed a bipartite graph of users and resources, we designed an extended graph representation to include the user's activities in the cloud servers. Specifically, we used a dynamic tripartite graph ~\cite{stahl1976genus} representation that consists of three disjoint sets of nodes (users, activities and cloud servers), where edges can only connect nodes from different sets, and not within the same set. In our representation, an edge cannot connect two users or two servers. The edges in this graph represent the connection between the users, actions and services over time.


Formally, we consider a tripartite dynamic graph $G$ with $n$ user nodes, $m$ server nodes, and $k$ action nodes denoted as: $V_u$, $V_s$, and $V_a$, respectively. 
The edges in $G$: $E \subset V_u \times V_a \times V_s \times \mathbf{N}^+$, represent connections between clients and cloud service via action. Therefore, for a user $u$ that accesses service $s$ using action $a$ two edges will be added to $G$ so that: $(u, a)$ and $(a, s) \in E$.
Additionally, an edge is assigned with time stamps and weight so that an edge $(v_i, v_j,t, w) \in E$ represents the number of connections $w$ between vertex $v_i$ and $v_j$ prior time $t$. For each time window, $T:[t_0,t]$,  let $A^{(T)} \in \mathbf{R}$ be the adjacency matrix of the graph in time $t$. 

\begin{figure}
  \includegraphics[width=\linewidth]{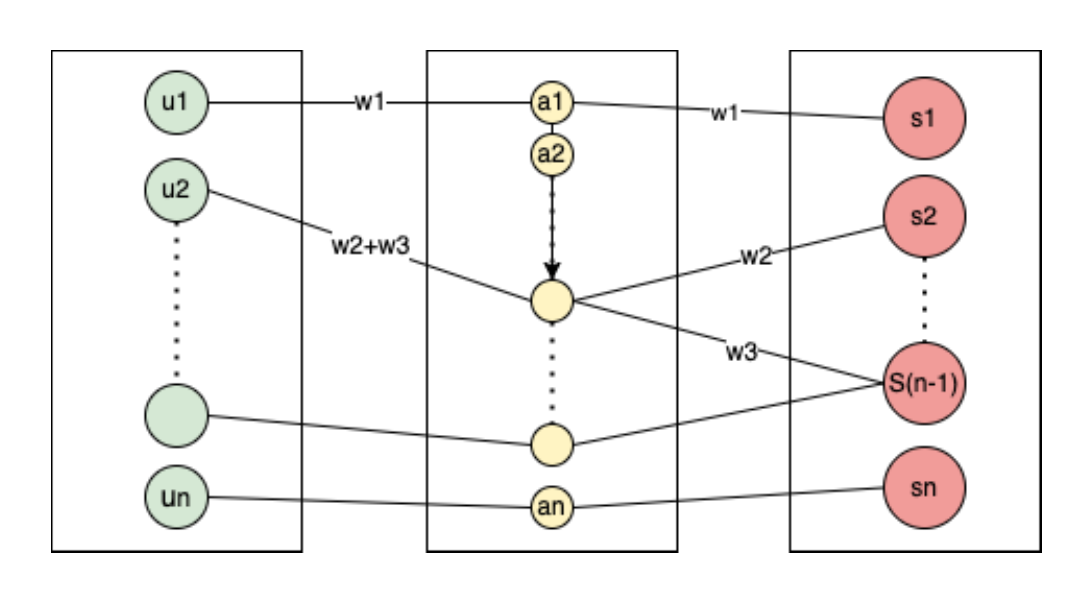}
  \caption{The proposed graph representation of the cloud services network. This is a tripartite graph where the green nodes represent the users, the yellow nodes represent the actions (Events), and the red nodes represent the cloud services. 
  Edge weights are defined with respect to the number of times the users approach the actions and the services. For example, user $u_1$ conducts action $a_1$ with respect to service $s_1$. On the other hand, user $u_2$ conducts $w_2+w_3$ actions, such that $w_2$ of those are with respect to service $s_2$ while the other $w_3$ are aimed towards service $S(n-1)$.} 
  \label{fig:Graphrep}
\end{figure}

Our approach sets each entry $A^{(T)}_{vu}$ to the number of times there has been a connection between vertices $u$ and $v$ prior to time $t$. Such graph representation aims to model the different graph components' activities throughout time.
Figure \ref{fig:Graphrep} illustrates an example of the dynamic tripartite graph representation.

\subsection{Node Embedding}
For the structural vector representation of each node in the graph, which also considers the weight of edges, we used the node2vec model ~\cite{grover2016node2vec}. This model is known for effectively capturing and preserving complex network structures in a low-dimensional space. Node2vec uses a random walk approach to generate the neighborhood of a node. This approach takes into account both immediate neighbors and far-off nodes information, creating robust vector embeddings. The node2vec algorithm is efficient and scalable which makes it suitable for large-scale systems representation such as cloud services systems.

\subsubsection{The Node2vec model embedding}
The node2vec model \cite{grover2016node2vec} is an unsupervised learning algorithm that generates continuous feature representations for nodes in a graph by preserving the network structure. Node2vec is an adaptation of the word2vec  algorithm ~\cite{mikolov2013efficient}, originally designed for natural language processing to create word embeddings. This model's algorithm consists of two main steps: generating random walks and learning node embeddings using a neural network.
\begin{itemize}
    \item \textbf{Generating random walks}:
For a given graph $G(V, E)$, where V is the set of nodes and E is the set of edges, the algorithm performs random walks starting from each node. The random walks aim to explore the local and global structure of the graph. In node2vec, the random walks are biased by two hyperparameters, $p$ and $q$, which control the exploration strategy.

 \item \textbf{Learning node embeddings using a neural network}:
Once the random walks are generated, a skip-gram neural network is used to learn the node embeddings. The objective is to maximize the probability of observing each of the node's neighbors for a given target node based on their co-occurrences in the random walks.
\end{itemize}

Mathematically, the optimization problem can be defined as follows:

Given a target node $u \in V$ with a set of neighbors $N(u)$, the goal is to maximize the following objective function:
\begin{equation}
    max_{f}= \sum_{u \in V }{ [-\log(Z_u) + \sum_{,n\in N(u)} f(n) \cdot f(u) ] }
\end{equation}
Where:  $Z_u$ is defined as 
a per-node partition that captures the similarity between the target vertex $u$ and the other vertices in $V$ under the function $f$. This function is mostly defined as the dot product between the function projection of two vertices. In the node2vec model, it is defined as follows:
\begin{equation}
Z_u= \sum_{v \in V}{\exp{f(u) \cdot 
 f(v)}}
\end{equation}

\subsubsection{Weighted random walks}

A random walk on a graph is a sequence of nodes that are visited by a random process that moves from one node to a neighboring node according to some predefined probabilities. A biased random walk is a type of random walk on a graph where the probability of transitioning from the current node to each neighboring node is not uniform but depends on some predefined rules or biases \cite{grover2016node2vec}.

The weighted random walk method extends the original random walks approach by incorporating edge weights into calculating walk biases during the random walk process. This approach uses the bias-weighted random walk based on two parameters: p and q. These parameters control the likelihood of returning to the previous node (p) or exploring new nodes (q) at each step of the walk. The random walk starts at a given node $v$ and performs a sequence of steps to generate a random walk of a certain length. At each step of the walk, the algorithm chooses one of the neighbor nodes of the current node $u$ to move to with this probability:


\begin{equation}
    P(v | u) = 
    \begin{cases}
    1/p & \text{if $(u, v) \in E $ }\\
    1 & \text{if $u = v$} \\
    1/q & \text{if $(u, v) \notin E$}\\
    \end{cases}
\end{equation}
where $E$ is the set of edges in the graph.

Each random walk sequence is used to construct a co-occurrence matrix that captures the pairwise relationships between nodes in the graph. The co-occurrence matrix is constructed by counting the number of times that each pair of nodes appears in the same random walk sequence.  Formally, a co-occurrence matrix M is a square matrix of size $n x n$, where $n$ is the number of nodes in the graph, and each entrance $M_{i, j}$ is defined as the number of times that nodes $i$ and $j$ appear together in the random walk sequence. This matrix is used as an input to a skip-gram ~\cite{mikolov2013efficient,mccormick2016word2vec} model to learn low-dimensional embeddings of the nodes in the graph.


Our approach is as follows: after processing the data as a graph of cloud entities, we apply the weighted Node2Vec algorithm to obtain node embeddings. This approach allows us to identify and visualize anomalies in the graph structure and gain insights into the underlying system's behavior.
The node2vec model provides a vector embedding for each node in the graph (users, activities and cloud servers). Denote the vectors $X_1,\ldots, X_m \in \mathbf{R_d}$ as embeddings of the users; our approach is to only use the user embeddings. Note that the vector embedding size of the Node2Vec model is a hyperparameter (typically in the range of 32 to 512). The choice of embedding size depends on various factors, such as the size of the graph, the complexity of the relationships between nodes, and the computational resources available for training the node. A common choice for the embedding size in Node2Vec is 128, which we adopted in the suggested solution architecture \cite{mikolov2013efficient}.


\subsection{Anomaly Score }
We dynamically assign an anomaly score for each user $u$ in a time $t$  as follows: Let $r$ be the cloud service last approached by $u$ and let $R=\{ u: (u,r, t^{'}), t^{'} \le t \} $, denote the set of users who have accessed service $r$ before time $t$. Let $X_u$ and $X_R$ be the vector embeddings for $u$ and $R$. 
The anomaly score for $u$ is determined using the Nearest-Neighbor (NN)~\cite{peterson2009k}  algorithm on the users embeddings in the following equation: 
\begin{equation}
    \text{Anomaly Score($u$)}=\text{NN($X_u, X_R$)} 
\end{equation}

In the nearest neighbor algorithm, the distance is quantified utilizing Euclidean similarity. By comparing the obtained anomaly score with a pre-established benchmark, one can determine if the recently added edge presents atypical properties.

For users that didn't access service $r$  before time $t$ the anomaly score will be $0$. This approach aids in monitoring new users accessing the service, as it enables the observation of score fluctuations over time and the identification of anomalous conduct by examining relative scores. 

\section{Dataset}
\label{sec:Dataset}
Amazon Web Services (AWS) is the largest cloud provider with a market share of 32 $\%$ \cite{MarketShare}. Hence, we decided to use AWS as our cloud provider for research. AWS provides the companies that use their cloud services with an activity tracking tool called CloudTrail. In order to test our proposed algorithm, we created a new benchmark dataset based on public CloudTrail records using Summit Route~\cite{cloudtrail}.
The Summit Route is a free training site for practicing attack techniques. This platform offers anonymized CloudTrail logs from "flaws.cloud".
Using this platform, we generated cloud services logs that include five common attacks embedded in simulated legitimate traffic. The legitimate traffic was simulated for each user in the cloud service by assigning it normal "day-to-day" tasks according to its role and access. 

Every enterprise that extensively utilizes cloud services from major providers such as AWS, Microsoft, and Google has the capability to monitor and track user activities within the cloud through user activity logs. In this research, we focused on leveraging the CloudTrail API calls recorder as can be seen in ~\cite{9932357} and ~\cite{routavaara2020security}, which serves as a foundational framework for detecting various attacks on AWS cloud services. CloudTrail, an AWS service specifically designed for auditing purposes, captures a comprehensive record of all user activities within an AWS account. These activities encompass a wide range of actions, including instance creation and termination, file uploads, and retrieval of database metadata through API calls which are recorded. Furthermore, CloudTrail provides valuable metadata pertaining to users and cloud services, such as timestamps, user IDs, service IDs, and status codes.
Our cloudtrail log includes 107116 API calls over 32 days. 
AWS has over 200 services. 
In our dataset, we tracked 79 different services.
Our dataset contains 5 attacks; four attacks were performed using one user (each) and one attack was performed using 4 users. In order to simulate the attacks we used a sandbox and downloaded the API calls performed by the users who performed the attack, and merged them into our dataset. In addition, we added previous legitimate actions to each user. Each attack occurs on a different day. However, all of the attacks follow several days of legitimate behavior to simulate cases when a legitimate user starts behaving in an abnormal way. To the best of our knowledge there isn't a labeled dataset available to the public.
 
\subsection {Attacks Simulation}
To simulate different kinds of user anomalies, we created five variations of common attacks:

    
\begin{itemize}  
\item {\bf Cryptojacking in a cloud account:} Cryptojacking~\cite{jayasinghe2020survey,tanana2020advanced} is a common attack where the attacker uses a user's cloud computing resources in order to create a distributed mining workforce. Once the attacker gains access to the user's resources, he launches new instances and installs malicious software on them. This manifests as launching and terminating instances in CloudTrail logs, which generates a massive irregular amount of RunInstances and  TerminateInstances (CloudTrail API calls) in our auditing service.
This attack may hold devastating results, such as increasing the cost of the cloud provider during this attack, halting current workloads and preventing the use of computing power due to quotas. 
\item {\bf Targeted billing attack:} In a targeted attack, the attacker's main goal is to take out small to medium-sized businesses by halting their development. Attacking the company's cloud services is the only way to perform this attack. This attack consists of launching/using expensive resources. Cloud providers charge on demand; for example, in a serverless database, using a lot of read/write methods can rack up the bills.
Similar to the previous attack, in this attack, an expensive instance type is launched to increase the cloud provider bills. Namely, an AWS instance type such as c7g.8xlarge costs around \$800 a month (depending on the region). Launching around 60 instances with a given type will result in a \$48000 bill; relaunching instances can rack up the bill even more. This type of attack can result in taking small businesses out of business and halting progress due to high costs for medium-sized businesses.
\item  

{\bf Lateral movement}:
In cloud terminology, the attacker obtains initial access to a user's cloud account. An attacker aims to elevate privileges, ultimately targeting the acquisition of sensitive information such as database copies, objects stored in storage services, etc. 

Our simulation implementation begins with exploiting a publicly accessible Log4j server to gain access to the first role. Roles are typically associated with instances to provide necessary permissions for application functionality. Once initial access is obtained, we leverage the user's permissions to discover the private key used to access the server. This discovery enables us to establish communication with a backend server that utilizes the same private key. Subsequently, we connect to the backend server, which activates a serverless lambda function with access to a secret manager. Through this mechanism, we retrieve a secret key that grants access to another user.
Consequently, the user can download the terraform state, which contains vital information for managing the infrastructure deployed through the popular Infrastructure-as-Code (IAC) tool, terraform. This information includes details regarding network peering, which can be activated by modifying a flag in the terraform code and deploying the changes. The user, facilitated by the lambda function, possesses the necessary permissions to perform these actions, thus establishing a connection between the backend server and the last server, which holds administrative privileges.

\item {\bf Monitor exploiting:}
In this scenario, the attacker starts by accessing a read-only role (usually for monitoring purposes). Since privilege escalation can be challenging or impossible in some cases, the attacker checks the network for vulnerabilities to exploit. In doing so, the attacker generates a lot of API calls to retrieve each piece of useful data and lay the ground for a future larger attack. The attacker scans most of the services in the cloud provider. Although information gathering can lead to a larger crisis, it often goes unnoticed. Because this isn't a "loud" attack, it doesn't rack up the cloud provider bill or mess with your functioning environments.

\item {\bf Targets specific services:} This attack targets specific services to exploit since the attacker doesn't have permission to modify or create resources. The attacker will target the existing resources, looking to find known services that may hold secret information or be able to further his permissions (usually, services that have role access to other services). The attacker doesn't modify or create resources, which sometimes makes it undetectable. In this case, the attacker targets misconfigured resources. An attack like this doesn't generate a lot of API calls, which makes it really hard to identify.
\end{itemize}

\section{Results}
In this section, we present the results of our study. This section describes the details of our framework in Section \ref{sec:Framework}, evaluates our method by comparing it to a baseline method in Section \ref{sec:baseline}, and finally, analyzes the results in Section\ref{sec:Experimental Results}.

\subsection{Framework}
\label{sec:Framework}

The initial stage of our experiment involved classifying the API calls into distinct groups. The categorization resulted in the formation of the subsequent classes.
\begin{table}
\begin{tabular}{|p{0.35\linewidth} | p{0.5\linewidth}|} 
 \hline
 Action & Description \\ [0.5ex] 
 \hline\hline
AssociateResources &  Connecting two resources with each other, for example, in AWS context associating a role with EC2 instance. \\
\hline
CreateObject & This has a variety of possible options, everything that includes creating a new resource, ranging from launching a new instance to creating a new bucket. \\
\hline
Delete & Each deletion of every object in AWS cloud, for example, terminating an instance \\
\hline
Disable objects & Make objects unavailable to work with until enabling them. \\
\hline
Download/UploadObjects & Downloading and uploading objects into AWS is a key component in multiple applications; as a result, we have given it a separate category. \\ 
\hline
EnableObjects & Making objects available to work with again after disabling them. \\
\hline
GetInfo & Get information about specific objects, receive an ip or OS of a currently running instance, information about a lambda function, etc. \\
\hline
GrantPermissions &  Granting permission to perform an action, whether it is for a role, attaching a role to a service or granting permissions to a specific user. \\ 
\hline
ListResources & List group of resources, list all running EC2 instances. \\
\hline
Login & Console login, role assuming and each call related to gaining permissions to perform an action. \\ 
\hline
Logout & Logout received its own category, since logging out multiple times from the same user can be a huge red flag. \\
\hline
ModifyExistingResource & Change specs of specific resources, change instance type (scale up/down), modify specific alert, modify RDS specs, etc. \\ 
\hline
RemovePermissions & Removing permissions if it is by deleting a policy, secret and access keys, etc. \\
\hline
SensitiveInfo & Each API call is directed at receiving sensitive information. This category is a bit complicated because we can only trust what is written in our API call and analyze it. Retrieving secret value from a secret manager, for example, always counts as SensitiveInfo. \\
 \hline
\end{tabular}
\caption{\label{actions}Action category table: our action categories and their description. }
\end{table}
Since there are over 3,000 different events in Cloudtrail, unique to each service, we decided to categorize the events. For example, retrieving sensitive information can be performed on a number of cloud services. Consequently, there isn't a need for a new node because they resemble the same action. Our categories and description can be seen in Table\ref{actions}.

Upon categorizing the actions, we divided our dataset into time frames. This was possible because our dataset contains a timestamp of each API call.
We represented each time frame as a graph, as mentioned in Section~\ref{sec:our_approach}. Afterwards, we activated the weighted node2vec algorithm ~\cite{grover2016node2vec},
in order to receive an embedding representing the user activity. 
The graph embedding encapsulates the users within the graphs as a vector, delineating the associations of each user with a specific action and, in turn, each action with a respective service.
We separated the graph into groups of users. Given an action $a$ and a user $u$, $X_a =  \{u \in X_a $ if $u$ performed $a$ during the given time frame\}, for each group ($X_a$) we graded each user ($u \in X_a $) within this group with an anomaly score by calculating the euclidean distance of the closest user in the same group.
Since we created the attacks, we could isolate the users who performed the attacks. We tested our method for each attack separately.

The method was tested on two different time gaps; long time gap and short time gap. In the case of a long time gap, we divided the dataset into 1-day additive time frames where each time frame contains the previous time frame plus one day starting from the earliest action's timestamp. 

\label{threshold}
Next, we describe the threshold calculation for bounding the anomaly score. We used the standard deviation and average of the anomaly score to determine which threshold is anomalous. Our threshold was calculated using a standard deviation score which is $1$ or $2$ standard deviations above the average using this equation: 
\begin{equation}
    \text{Threshold}_n = \mu + n\sigma
\end{equation}
Where $ \mu$ represents the mean of the data, and $\sigma$ represents the standard deviation and $n=\{1,2\}$.
This threshold generates the best results in terms of the lowest percentage of false positives while not damaging the detection rate of the attacks. This was also suggested as a robust threshold in \cite{reimann2005background, shahid2022detecting} 





\subsection{Baseline}
\label{sec:baseline}
 To evaluate the effectiveness of our method, it is essential to compare it against a reliable baseline. For this purpose, we used the ~\cite {modell2021graph} approach as our comparative standard. This approach utilizes a dataset of users accessing a range of shared servers. They partitioned their dataset into time-specific segments, much like our own methodology. Their graphical representation consisted of two primary nodes, the user nodes and server nodes, further enriching the data analysis process.
In the baseline method,  each time frame is represented as a graph such that each user and each server is a node. An edge consisted between the user and server if the user accessed the server during the given time frame with no regard to the number of actions performed by the user. Instead of the node2vec embedding method, the baseline method uses spectral embedding for the anomaly score distance calculation. The anomaly score algorithm divides the graph into groups. Each server has its own group of users that access the server in the given time frame. The anomaly score was given to a user in a server group in two steps: if the user has accessed the server in previous time frames, the anomaly score is 0, otherwise, the user was assigned an anomaly score by calculating the closest neighbor in that server group.

\subsection{Experimental Results}
\label{sec:Experimental Results}
\begin{figure}
 \includegraphics[width=\linewidth]{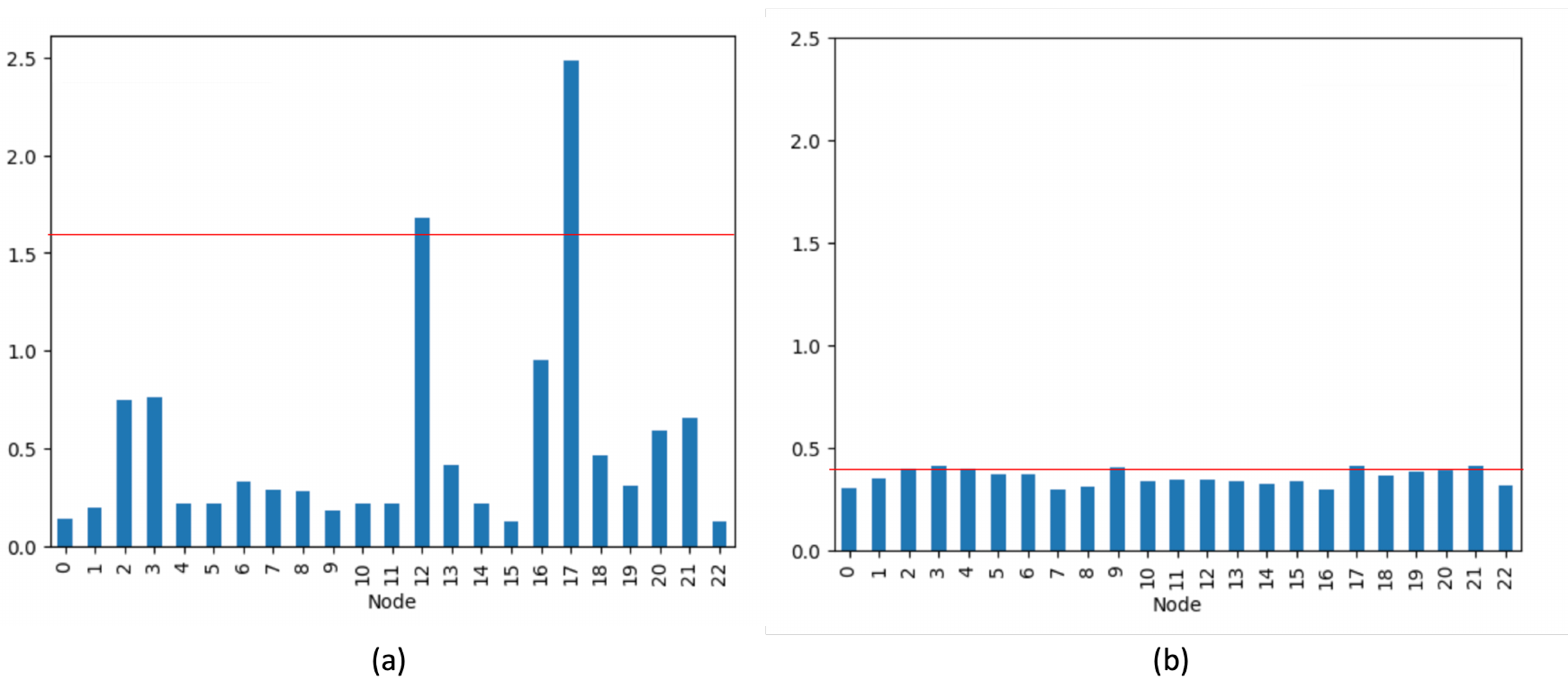}
  \caption{ (a) Targeted Attack:  CS-GAD result. The red line represents the threshold line. (b) Targeted Attack: baseline result. The red line represents the threshold line.}
  \label{fig:anomalyScore}
\end{figure}

Our experiment determined the system's merit using the false positive rate as a crucial metric. Indeed, the importance of detecting an attack cannot be overstated. However, maintaining a low false positive rate is arguably of equal significance. A high rate of false positives can breed complacency, leading to a lessened seriousness attributed to alerts. This could culminate in real attacks being overlooked and potentially result in delayed detection or an attack going completely unnoticed. Table \ref{results} shows the false positive rate using our approach and the baseline method. Figure \ref{fig:anomalyScore} demonstrates the threshold detection for both the baseline and our method. \newline
\begin{table}[H]
\begin{tabular}{| p{0.2\linewidth} | p{0.15\linewidth} | p{0.15\linewidth} | p{0.15\linewidth} | p{0.15\linewidth}|} 
\hline
Attack Name &\textbf{CS-GAD}-FPR AVG + STD * 1 & \textbf{CS-GAD}-FPR AVG + STD * 2 & \textbf{Baseline}-FPR AVG + STD * 1  & \textbf{Baseline}-FPR AVG + STD * 2  \\ [0.5ex]
\hline\hline
Crypto-jacking & 12\% & 4\% & 25\% & 10\% \\
\hline
Billing attacks & 4\% & 4\% & 28\% & 13\% \\ 
\hline
Lateral movement & 25\% & 8\% & 32\% & 16\% \\ 
\hline
Vulnerability scan & 14\% & 5\% & 23\% & 9\% \\
\hline
Targeted vulnerabilities in services & 20\% & 10\% & 27\% & 12\% \\
\hline

\end{tabular}
\caption{\label{results} False positive rate (FPR) for each attack with different standard deviations (STD) from the average (AVG).}
\end{table}

Table \ref{results} shows that our false positive rate is lower than the baseline algorithm. This table also shows that the 2 STD distance from the average, as described in Section \ref{threshold} is the more accurate threshold for our data-set. As can be seen in our results, this has the lowest false positive rate without decreasing the detection rate of the users who performed each attack. However, for different data sets, some other thresholds may fit better.

\begin{table}[H]

\centering
\begin{tabular}{|p{0.3\linewidth} | p{0.2\linewidth}|p{0.25\linewidth}|} 
\hline
\textbf{} & \textbf{CS-GAD} & \textbf{Baseline} \\ 
\hline\hline
Accuracy & 0.86 & 0.80 \\
\hline
Precision & 0.66 & 0.55 \\
\hline
Recall & 0.85 & 0.71 \\
\hline
F1-Score & 0.75 & 0.62 \\
\hline
\end{tabular}
\caption{\label{Measurements}  Baseline and CD-GAD Accuracy.}
\end{table}

Table \ref{Measurements} compares our accuracy, precision, recall and F1-score of the algorithms.

\subsubsection{User anomaly score distribution} \label{sec:anomaly score}
In this section, we will look closer at user score distributions of the CS-GAD and baseline methods and understand the importance of a wider spread of user anomaly scores, meaning larger variance. Comparing and analyzing our proposed method against the baseline is done by comparing the variance of scores across different attacks. Since our goal is to ensure we can identify anomalies, we confirm that a user with abnormal behavior would have a score with high variance compared to their peers. Table \ref{Variance} presents the variance of both methods for each attack.
\begin{table}[H]

\centering
\begin{tabular}{|p{0.3\linewidth} | p{0.2\linewidth}|p{0.25\linewidth}|} 
\hline
\textbf{Attack Name} & \textbf{Variance Baseline} & \textbf{Variance CS-GAD} \\ 
\hline\hline
Crypto-jacking & 0.0084 & 0.3974 \\
\hline
Billing attacks & 0.0081 & 0.5207 \\
\hline
Lateral movement & 0.0055 & 0.3044 \\
\hline
Vulnerability scan & 0.0031 & 0.3295 \\
\hline
Targeted vulnerabilities in services & 0.0072 & 0.3192 \\
\hline

\end{tabular}
\caption{\label{Variance} Variance table - The distribution variance compared to the baseline.}
\end{table}

Compared to the baseline, we can see that our variance is higher for all attacks. This can be explained as follows: In our approach, graph embedding enhances the distance between malicious and legitimate users. Legitimate users are projected closer to each other, while malicious users are far from every other user, whether it is malicious or not. Because of this, we recognize, differentiate, and group legitimate and malicious users better. In the baseline, legitimate and malicious users have a closer anomaly score and can be more easily missed.

\subsubsection{Adjusting time frames windows}

Next, we discuss the trade-off between choosing a large time window and decreasing accuracy. 
Our proposed method is sensitive to some external parameters that define its precision level and, consequently, the run time. One significant parameter is the choice of time window size for data partitioning, as it determines the resolution at which we observe the dynamic network of cloud calls. Naturally, decreasing resolution enhances the accuracy of the method. In our case, it is caused by two reasons: Firstly, the larger our time window, the smaller the relative score of each user compared to their peers. This is due to the increased variance between the scores, as explained in Section \ref{sec:anomaly score}.
Secondly, the smaller the time window, the earlier we can identify attacks, simply because we perform more anomaly score checks within the same time window. 

\begin{figure*}[h]

    \centering
        \begin{subfigure}[t]{0.5\textwidth}
        \label{First}
        \centering
        \includegraphics[height=2in,width=3in]{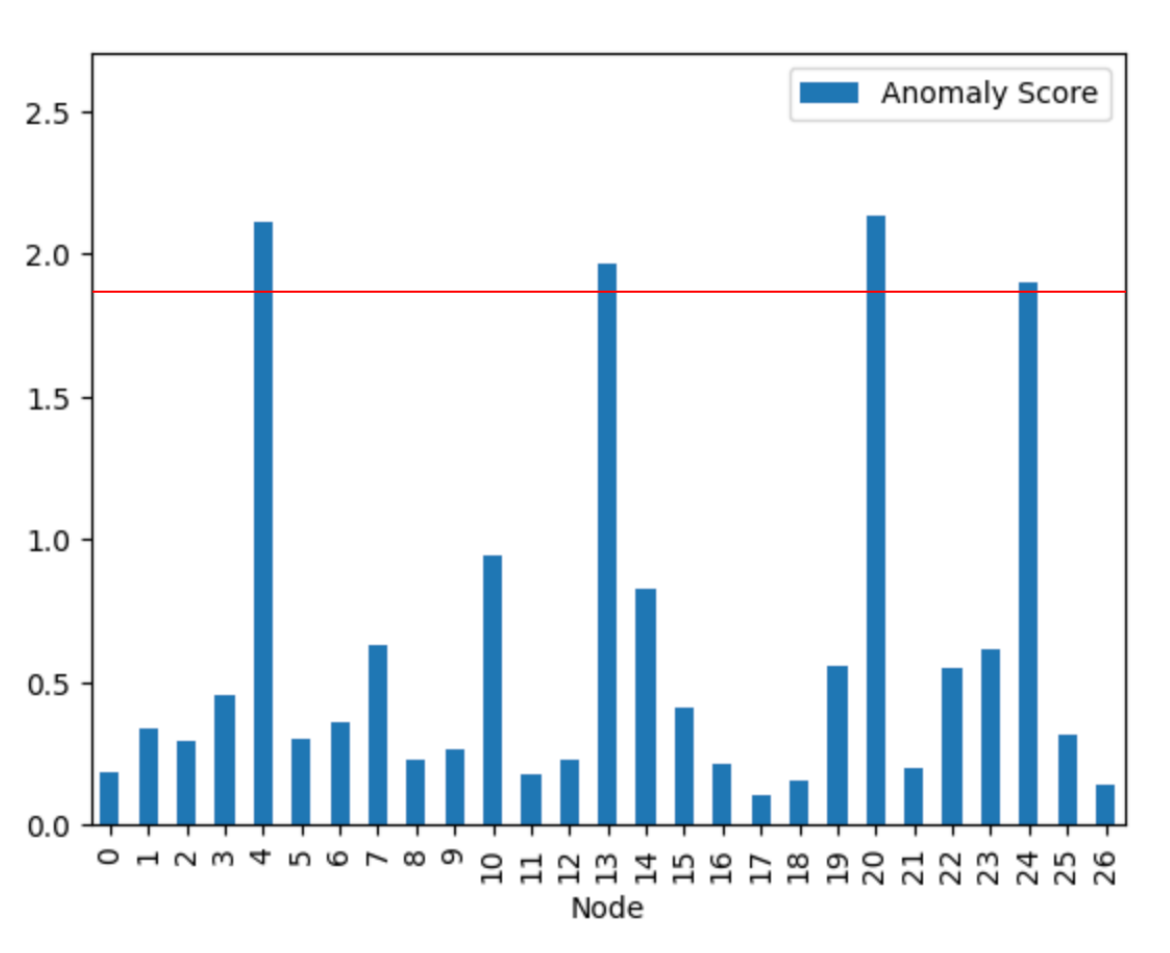}
     \caption{Results divided to one-hour time-window}
    \end{subfigure}
    \begin{subfigure}[t]{0.4\textwidth}
    \label{Second}
        \centering
        \includegraphics[height=2in,width=3in]{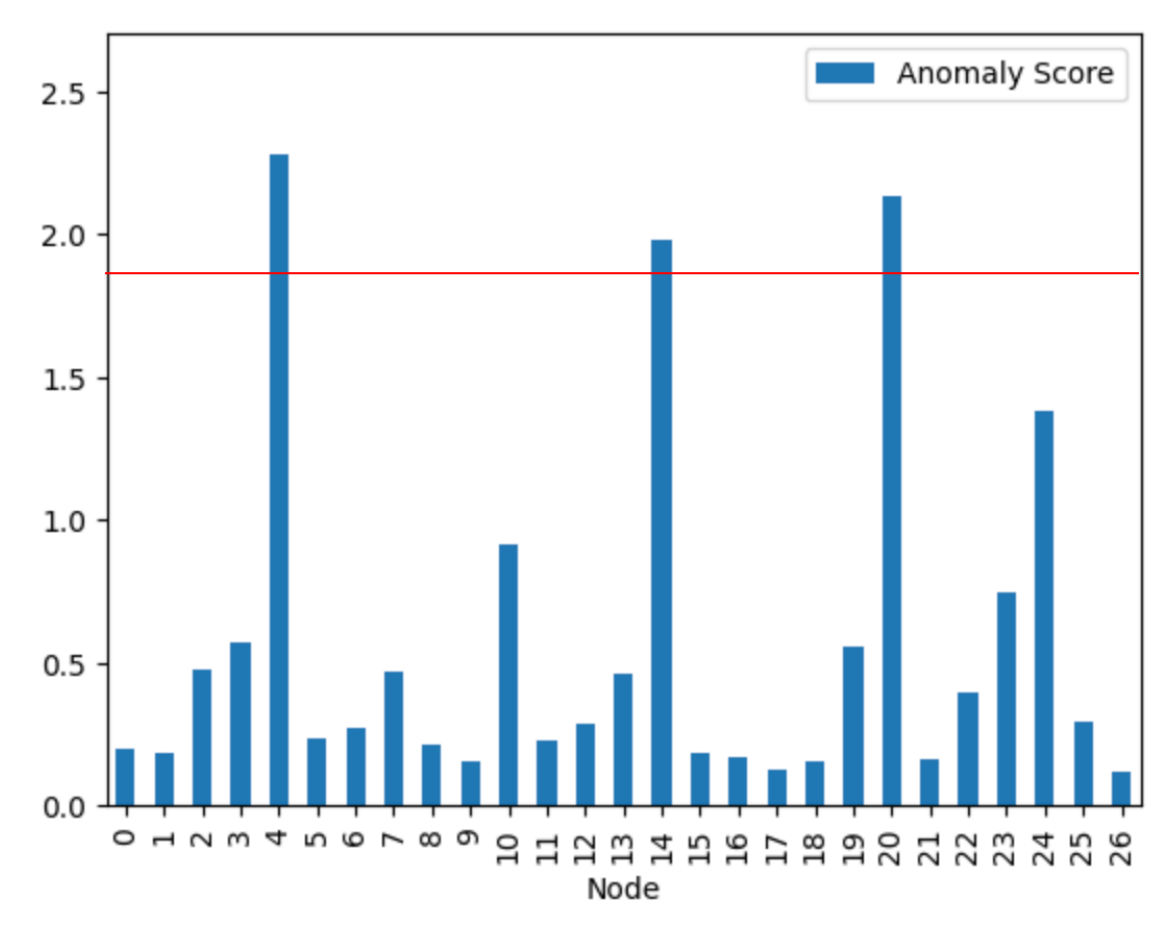}
       \caption{Results divided to one-day time-window} 
    \end{subfigure}~
    \caption{Our results in lateral movement attack compared in different time-window sizes. Figure (a) shows the algorithm detects two users out of the five who performed that attack, while Figure \ref{comparison}(b) shows that our algorithm detects only one user who performed the attack
    }
    \label{comparison}
\end{figure*}

However, reducing the time window size comes at a cost in terms of higher runtime, as we compute the graph embedding more frequently, which is a computationally expensive process. Therefore, for each dataset, it is beneficial to assess the viability of different time windows in terms of runtime and accuracy and determine the threshold between them. 
Next, we will present our analysis of the selection of different time window sizes for the dataset presented here.
In our experiments, we used two different time-windows: one-day and one-hour time windows to evaluate our method as explained in Section \ref{sec:Framework}.  Figure \ref{comparison} shows the results of our method using different time windows in the lateral movement attack. 
As shown in Figure \ref{comparison}(a), user 24 was overlooked when we used a larger time window. This was primarily due to their involvement in a lateral movement attack, a strategy in which multiple users are engaged over time and their behaviors increasingly resemble each other, as they access similar resources. In this example, our method detected the initial attackers but failed to recognize the remaining ones when employing a large time window (one day). Conversely, an additional attacker was identified when we utilized a smaller time window (one hour) in Figure \ref{comparison}(b).

Deciding the time window size is key to increasing the detection rate. After conducting multiple tests, we have found that using a one-hour time window size is optimal for detecting anomalies and accurately differentiating between malicious and legitimate users on datasets of our scale.



\section{Conclusion and Future Work}
Cloud-based information management has become a prevalent field in recent years, with many companies leveraging cloud services for their daily data storage and computation needs. In light of this, a problem of increasing external attacks on these companies has emerged, intending to cause harm or steal cloud computing time for various purposes. These attackers exploit the vulnerabilities of individual cloud network users who have a range of access permissions. Such attacks often go unnoticed, as many user activity monitors don't typically reveal any deviations in their permissions.

This study presents a method that uses a graph representation of user activities in cloud services over time. This method employs a GNN algorithm for graph embedding of these activities and assigns an anomaly score to each user. We have demonstrated that this method is robust in identifying different types of attacks and yields relatively low false positive rates compared to a basic method by $2$ - $9$ percent. Furthermore, we have constructed a benchmark of cloud network traffic that simulates network attacks and normal user behavior. This benchmark is open to all and serves as a valuable resource for further research and development in this area.

In the future, we intend to evaluate our algorithm on datasets of a larger scale to test its efficacy and scalability. Moreover, we aim to refine our approach by taking into account a range of additional factors, including multiple resource creation via a single API call and the consideration of status codes. Looking ahead, we aim to also build preventive actions within the cloud system. This could mean limiting user permissions if we spot something odd, effectively putting that user on hold until we can be sure their activity is safe. This way, we're not only identifying problems but taking active steps to stop them, helping to make the cloud environment more secure.

\section*{Acknowledgment}
This work was supported by the Ariel Cyber Innovation Center in conjunction with the Israel National Cyber Directorate in the Prime Minister's Office.

\bibliographystyle{unsrtnat}  
\bibliography{references}

\end{document}